\documentclass[aps,prd,twocolumn,groupedaddress,eqsecnum,amssymb,showpacs,
nofootinbib]{revtex4}
\usepackage{graphicx,mathptm,bm,times,epsfig}

\newcommand{\be}{\begin{equation}}
\newcommand{\ee}{\end{equation}}
\newcommand{\bea}{\begin{eqnarray}}
\newcommand{\eea}{\end{eqnarray}}

\begin{document}

\title{'Tilting' the Universe with the Landscape Multiverse: The 'Dark' Flow}

\author{L.~Mersini-Houghton}
\affiliation{Department of Physics and Astronomy, UNC-Chapel Hill, NC, 27599-3255, USA}
\email[]{mersini@physics.unc.edu}
\author{R.~Holman}
\email[]{rh4a@andrew.cmu.edu}
\affiliation{Department of Physics, Carnegie Mellon University, Pittsburgh PA 15213, USA}

\date{\today}

\begin{abstract}
The theory for the selection of the initial state of the universe from the landscape multiverse predicts superhorizon inhomogeneities induced by nonlocal entanglement of our Hubble volume with modes and domains beyond the horizon. Here we show these naturally give rise to a bulk flow with correlation length of order horizon size. The modification to the gravitational potential has a characteristic scale $L_{1} \simeq 10^{3} H^{-1}$, and it originates from the preinflationary remnants of the landscape. The 'tilt' in the potential induces power to the lowest CMB multipoles, with the dominant contribution being the dipole and next, the quadrupole. The induced multipoles $l \le 2$ are aligned with an axis normal to their alignment plane being oriented along the preferred frame determined by the dipole. The preferred direction is displayed by the velocity field of the bulk flow relative to the expansion frame of the universe. The parameters are tightly constrained thus the derived modifications lead to robust predictions for testing our theory. The  'dark' flow was recently discovered by Kashlinsky et al. to be about $700 km/s$ which seems in good agreement with our predictions for the induced dipole of order $3 \mu K$. Placed in this context, the discovery of the bulk flow by Kashlinsky et al. becomes even more interesting as it may provide a probe of the preinflationary physics and a window onto the landscape multiverse.

\end{abstract}


\maketitle

\section{Introduction}
\label{sec:intro}

The birth of our universe is one of the most important mysteries left in our understanding of nature. It has also been one of the most notoriously difficult problems to address. The program we developed over the last few years \cite{lauraarch,laurarich,avatars1,avatars2, laurareview} advocates a new approach to this central problem. Our investigation of the selection of the initial conditions of the universe differs radically from the previous attempts. This program is based on the point of view that: the initial system must be studied as a {\it non-equilibrium} event; the quantum dynamics of the gravitational degrees of freedom needs to be incorporated; and, our physical theories have to be extended to a multiverse framework.

Here we show how the observed 'dark' flow is naturally predicted in this theory to be a remnant of the birth of the universe from the landscape multiverse.
Our theory is based on the sole assumption that quantum mechanics is valid at all energy scales. The calculation is carried out by proposing to allow the wavefunction of the universe to propagate on the landscape multiverse \cite{lauraarch}. Its cornerstone achievement is that it provides a sensible answer to the question: 'why did our universe start from such an extremely ordered state?' \cite{laurarich, laurareview}. The dynamics of the wavefunction leads to a {\it superselection rule} that selects {\it 'survivor'} universes to be those born at energies high enough, (ordered states), that survive the backreaction of matter modes, and {\it 'terminal'} universes to be those initial states that start at sufficiently low energies that can not survive the collapse induced by the backreaction of matter modes onto the initial patch. The working model we used \cite{laurarich} for demonstrating the application of our proposal, was the landscape of string theory since string theory is the leading candidate for quantum gravity at present. 

In order to derive testable predictions for this theory, in 2006 we embarked upon an investigation of remnants in present day observables, of the preinflationary dynamics that led to the birth of our universe from the landscape multiverse. The handle we used to track down the preinflationary remnants of the matter backreaction and nonlocal entanglement with modes beyond the Hubble volume, was based on the unitary evolution of the universe. According to unitarity, information about the preinflationary remnants of superhorizon sized entanglement and non-equilibrium dynamics  of this system on the landscape can not be lost. Such investigation \cite{avatars1,avatars2} led to a series of predictions for traces on CMB and LSS of remnants of the preinflationary dynamics of our state on the landsacape. Two of these predictions, namely: a giant void at $z\simeq 1$ and size about $200$ Mpc, as well as a suppressed amplitude of the CMB fluctuation,induced by the superhorizon non-inflationary inhomogeneities, $\sigma_8 \simeq 0.8$, were succesfully tested recently by radio measurements \cite{rudnick} and,
WMAP -{\it 5-year} data \cite{wmap5}. A by-product of this analysis were the predictions for another void near the quadrupole, and of the alignment of the lowest CMB multipoles. These remnants originate from the same mechanism, the 'tilting' of the potential induced by the superhorizon entanglement \cite{avatars1,avatars2,laurareview}.

The scope of this Letter is to show that the recent observation of the 'dark' flow \cite{Kash} tests the third prediction we made in \cite{avatars2} namely, the existence of superhorizon sized non-Gaussian and noninflationary inhomogeneities that 
arise from the backreaction of superhorizon matter modes and nonlocal entanglement of our universe with all other patches and massive modes beyond the horizon.
Considering that the characteristic length of the noninflationary inhomogeneities $L_1$ is $10^3$ times larger than the horizon $r_H$, then their main contribution is to add power to the lowest CMB multipoles. For clarity we explicitly calculated the contribution of these superhorizon inhomogeneities to the quadrupole in \cite{avatars2, avatars1}. We showed they produce an induced quadrupole in the CMB, \footnote{ see Eq.(3.5) in \cite{avatars2} }, which is added to the inflationary one.  The noninflationary modification to the gravitational potential by these superhorizon inhomogeneities was also derived, (Eq. (3.4) in \cite{avatars2} ), and as shown there \cite{avatars1,avatars2,laurarich} this derivation makes no allowance for phenomenology, since its parameters are tightly constrained. We derived the {\it quantum coherence length} $L_1$  \cite{avatars1}, a length that  describes the distance at which the quantum interference effects between our Hubble volume and modes beyond the horizon become significant and of order unity relative to the inflationary density perturbations. For example, for $GUT$ scale inflation, the expression for $L_1$ calculated in \cite{avatars1} gives a coherence length of order $L_1 \simeq 900 r_H$ with $r_H$ the horizon radius. The cumulative effect of the superhorizon noninflationary inhomogeneities thus has a characteristic wavenumber  and scale $k_1 \simeq 1/L_1$.  Besides the quadrupole, the nonlocal modification to the gravitational potential by these modes also induce a dipole in the CMB. The dipole can be estimated within our model of the landscape multiverse as shown in Sec.3. The rest frame determined by the noninflationary dipole does not coincide with the expansion frame, rather it induces relative motion with respect to the expansion of the universe. The noninflationary dipole has the same amplitude as the one derived for the quadrupole in \cite{avatars2} but it is a factor of $O(L_{1}/r_{H})$ larger than the quadrupole. We estimate the bulk flow arising from the induced dipole predicted here \cite{avatars1,avatars2} and compare its value to current data. 

Kashlinsky {\it et al} \cite{Kash} have recently claimed a stunning discovery: they observe
just such coherent cosmological bulk flow of matter aligned along the dipole, on scales out to at least $300 h^{-1}$ Mpc, and with amplitudes of order $600-1000\ {\rm km \slash sec}$ by using the kinetic Sunay'ev-Zel'dovich effect (KSZ). Such a directional flow relative to the expansion of the universe frame, derived by comparison with the cosmic $X-$ray background, provides evidence that the induced dipole in the CMB is of noninflationary origin. They see no sign of convergence of such a bulk flow which indicates that the coherence length is at least horizon size. Thus the name: 'the dark flow'. Since the correlation length of such a flow is horizon size then the noninflationary perturbation that gives rise to the dipole has to be induced by superhorizon inhomogeneities. But, considering that the flow provides a new preferred frame relative to the expansion then these inhomogeneities can not be of inflationary origin. 

These results are extremely difficult to accomodate within the paradigm of gravitational instability in the concordance $\Lambda$CDM cosmology. In this cosmology, the peculiar bulk velocity would decrease as $V_{\rm rms} \propto r^{-1}$ on a scale $r$ and should be extremely small on the scales probed in Ref.\cite{Kash}. Besides, the observed dipole in an inflationary universe is simply kinematic, i.e. due to our local relative motion with respect to the galactic plane. The intrinsic CMB dipole is canceled out by the gravitational potential term, therefore unobservable. In other words in concordance cosmology, the rest frame defined by the expansion of the universe coincides with the CMB rest frame. The directional motion of clusters aligned with the dipole and with horizon size correlation length, i.e. the 'dark flow' observed by \cite{Kash}, provides a second frame relative to the expansion of the universe frame and indicates that corrections to the gravitational potential for the induced dipole must be due to superhorizon sized noninflationary inhomogeneities.

The modification to the gravitational potential in the universe, produced by remnants of the {\em quantum nonlocal entanglement}, and the subsequent quadrupole derived in \cite{avatars2},  are reviewed in Sec.2. The corresponding noninflationary dipole induced by the 'tilted' potential is discussed in Sec.3. The predicted imprints trace out the non-local entanglement effects among different horizon regions and modes beyond our horizon, that started in different sectors of the landscape and interfere with one another to induce decoherence. The derivation of the decoherence effects on the wave function of the Universe was done via the usual mechanism of tracing out superhorizon fluctuations to construct an effective theory that describes inflationary physics within the given horizon.

Our contention, then, is that these observations of bulk flow can be construed as evidence for the birth of the universe from the landscape multiverse imprinted on the superhorizon sized nonlocal quantum entanglement between our horizon patch and others that began from the landscape. When we calculate the size of the induced dipole in our theory and convert it into a bulk velocity dispersion, we will see that for the constrained values of our parameters we arrive at a velocity dispersion of order $\sim 670\ {\rm km\slash sec}$, remarkably close to the observed value of $\sim 700\ {\rm km\slash sec}$ !

\section{Review: Birth of the Universe from the Landscape Multiverse}
\label{sec:model}

We will only sketch out the model we use here so as to set the stage for our explanation of the dipole anisotropy and refer the reader to Refs.\cite{landscape,laurareview,laurarich} for more details. 

Think of the panoply of all possible ground states for the true theory of everything as a lattice of potential wells. We proposed in \cite{lauraarch} to let the wavefunction of the universe to propagate on the landscape of these potential wells. The superspace for the wavefunction is determined by the values of the various moduli fields needed to specify the vacua, as well as on the gravitational variables which is the scale factor $a[t]$ of the $3-$geometries, \cite{deWitt}; these variables form the configuration space for the wavefunctions. If we specialize to the string landscape, we can separate the lattice sites into those that describe supersymmetric vacua and those in which supersymmetry (SUSY) is broken. The wavefunction follows a Bloch wave pattern on the SUSY sector thus it is not of interest to cosmology since decoherence is hard to achieve. But in the broken SUSY sector, the vacuum energies are randomly distributed, as shown by Douglas and Denef \cite{landscape} within a certain range given by the string scale. We treat it as a {\em disordered} lattice. The strength of the disorder is of order the string scale or the Planck scale, since the fundamental parameter of the theory determines the range of vacua energy distribution. 

Disordered lattices have the interesting possibility of Anderson localization \cite{anderson} as proposed in \cite{lauraarch}. This allows for the localization of the wavefunction around each one of the broken SUSY lattice sites, which in turn allows us to think of each lattice site as the potential starting point of new ``baby universe'' in the sense that different horizon regions could have started from different vacua characterized by the physics of the given vacuum site. For this reason, we proposed in \cite{laurareview,lauraarch} to view the {\it landscape of string theory as the physical phase space for the initial states} and view the ensemble of all wavefunction solutions born out of this landscape as the {\it physical multiverse} predicted by the landscape of string theory. However, since the full wave function of the Universe contains all that information, the physics in any given site must be entangled with that of {\em all} other lattice sites. Each initial wavepacket in the family of solutions for the wavefunction of the universe contains matter and vacuum energy. Decoherence among such solutions is induced by the backreaction of massive perturbation around the vacua site where the wavepacket is localized. With decoherence taken into account, the quantum dynamical evolution of matter and gravitational degrees of freedom of the wavepacket leads to the following superselection rule: the initial states that contain a vacuum energy high enough to overcome collapse from the backreaction of matter modes result in 'survivor' universes, whereby the quantum to classical transition for this universe is succesful. Such initial domains can grow to physically large classical universes. But initial wavepackets that start at low energies can not overcome the collapse induced by the backreaction of matter and thus lead to 'terminal' universes since these states do not survive the transition to the classical phase and can not grow. The latter are wiped out of the phase space of initial conditions, i.e. from the ensemble of the initial states which construes the multiverse. One of the 'survivor' universes leads to the birth of our universe, \cite{laurarich}. The subset of the original ensemble selected as 'survivor' universes comprises the present multiverse. 

The initial states that led to the birth of physical universes follow a unitary evolution. It is known that a mixed state can not evolve into a pure state under a unitary evolution \cite{avatars1,avatars2}, a fact that ensures that traces of the initial entaglement among different patches and of the backreaction of superhorizon matter modes onto our patch, do survive in the present observable sky. The cumulative effect of the superhorizon nonlocal entanglement among the various domains and fields has a number of physical consequences. First, in order to understand the physics inside a given horizon, superhorizon fluctuations present in the wave function of the universe must be traced out, generating a reduced density matrix for the relevant variables. As usual in a reduced density matrix, the off-diagonal coherences measure how entangled the system is, and in particular, they give a measure of the scale $L_1$ on which these cumulative entanglement effects are large. Another way to understand this scale, denoted by $L_1$ is to think of it as the minimum distance we need to go far out of the horizon upon which quantum interference effects with other domains and superhorizon modes become sufficiently significant for revealing the quantum nature of the universe. Equivalently, $L_1$ is the scale at which quantum interference corrections becomes of the same order as the classical curvature inhomogeneities.

We showed in \cite{avatars1,avatars2} how the tracing out of superhorizon modes gives rise to corrections to the Friedmann equation for the scalar field that drives inflation. These corrections are not arbitrary, but are in fact related to parameters specifying the broken supersymmetry sector of the landscape. We have computed these corrections in previous work, and list them below. The modified Friedmann equation for the inflaton field with a potential $V(\phi)$ on our domain, being embedded in the multiverse, now becomes:
\begin{eqnarray}
\label{eq:modfried}
H^2 &=& \frac{1}{3 M_{\rm Pl}^2} \left[V(\phi) + \frac{1}{2} \left(\frac{V(\phi)}{3 M_{\rm Pl}^2}\right)^2 F(b, V(\phi))\right]\nonumber\\
&\equiv & \frac{V_{\rm eff}(\phi)}{3 M_{\rm Pl}^2}\nonumber\\
F(b,V(\phi)) &=& \frac{3}{2} \left(2+\frac{m^2M^2_{\rm P}}{V(\phi)}\right)\log \left( \frac{b^2 M_{\rm P}^2}{V(\phi)}\right)\nonumber \\
&-&\frac{1}{2} \left(1+\frac{m^2}{b^2}\right) \exp\left(-3\frac{b^2
M_{\rm P}^2}{V(\phi)}\right).
\end{eqnarray}
Here we have taken $8\pi G_N =  M_{\rm Pl}^2$, $m^2 = V^{\prime \prime}(\phi)$ and $b$ is the SUSY breaking scale, which in our picture also corresponds to the width of the Anderson localized wavepackets centered about a vacua site of the landscape. We can understand the nonlocal modification term, $F(b,V)$, to the Friedmann equation, by noting that the entanglement and backreaction of other patches and modes on our initial state induces a shift in the energy of the wavepacket of our universe, correspnding to a deviation from its classical path in the phase space. This information is contained in the phase of the perturbed  wavepacket, (see Eq.(3.3-3.8) in \cite{avatars1} for the derivation of this modification).

As mentioned above, the off-diagonal elements of the reduced density matrix give us information about the amount of entanglement present between vacua. In particular, the characteristic length scale $L_1$ of the quantum interference, which contains the information about entanglement with all the different domains and superhorison modes, was derived in \cite{avatars1} in terms of the parameters of the model:
\begin{equation}
\label{eq:coherencelength}
L_1^2 =
\frac{a}{H}  \left[ \left(\frac{m^2}{3 H}+H\right)\ln\frac{b}{H}-\frac{m^2 H}{6}\left(\frac{1}{b^2}-\frac{1}{H^2}\right) \right].
\end{equation}

As can be seen by these expressions and their derivation in \cite{avatars1,avatars2}, there is no room allowed for phenomenology.

\section{'Dark' Flow, Superhorizon Entanglement and the Induced Dipole}
\label{eq: bulk}

We reviewed how tracable remnants for testing our theory for the birth of the universe, are induced by the superhorizon sized inhomogeneities with a characteristis scale given by $L_1$ that arise from the nonlocal entanglement of our domain with modes and patches beyond the horizon. These noninflationary superhorizon inhomogeneities are added to the inflationary CMB, leading to a modification of the Newtonian potential $\Phi$  Eq.\ref{eq:tilt}, derived in \cite{avatars1,avatars2}. The nonlocal modification to the gravitational potential of the universe is relevant here because the 'tilt' induced by the modification in the potential drives the bulk flow as we now demonstrate. The 'tilted' potential with the subsequent residual (noninflationary) quadrupole induced by it, were calculated in Ref. \cite{avatars1,avatars2} to be:

\begin{equation}
\label{eq:tilt}
\Phi =\Phi^0 + \delta\Phi \simeq \Phi^0 \left [1 + \frac{V(\phi) F(b,V)}{3M_{pl}^2} (\frac{r}{L_1}) \right]
\end{equation}

where $\Phi^0$ is the inflationary, standard gravitational potential before modifications $\delta\Phi$ are included, and the induced quadrupole anisotropy resulting from this potential,

\begin{equation}
\label{eq:indquad}
\left . \frac{\Delta T}{T}\right |_{\rm quad} \simeq 0.5 \left (\frac{r_H}{L_1}\right )^2 \left(\frac{V(\phi) F(b, V(\phi))}{18 M_{\rm Pl}^4}\right ).
\end{equation}

From the 'tilt' modification term $\delta\Phi$ in the potential $\Phi$, Eqs.\ref{eq:tilt}, \ref{eq:indquad}, it is straighforward to derive the expression for the induced dipole, 

\begin{equation}
\label{eq:inddip}
\left .  \frac{\Delta T}{T}\right |_{\rm dip} \simeq \frac{4\pi}{15} \left (\frac{r_H}{L_1}\right ) \left(\frac{V(\phi) F(b, V(\phi))}{18 M_{\rm Pl}^4} \right ).
\end{equation}

The nonlocal effects only modify the largest scales. We can see this from Eq.~(\ref{eq:tilt}), since the ratio $r\slash L_1$  encoded in $\delta\Phi$, with $L_1 \gg r_H$,  can significantly affect power only to the CMB multipoles of the largest scales, i.e. for $r\simeq r_H =H^{-1}$. In fact as can be seen from our dipole expression, even the amplitude of the induced quadrupole $l=2$ is already suppressed by a factor of order $O({r_H}\slash{L_1})$ as compared to the dipole. This is equivalent to the fact that since the distance scale for our noninflationary perturbations is $L_1 \simeq {1}\slash{k_1}$ then $\vec{k}\cdot \vec{x} \ll 1$. Thus, since $\delta \rho\ \exp(-i\vec{k}\cdot \vec{x}) \simeq \delta \rho [1-i\vec{k}\cdot \vec{x} -\frac{(\vec{k}\cdot \vec{x})^2}{2} +...]$, the multipole decomposition of the density perturbations will be dominated by the lowest multipoles. We perceive such a long wavelength perturbation as a linear gradient field because the induced multipoles are dominated by the first correction term, $\vec{k}\cdot \vec{x}$ {\it i.e.} the induced dipole, with the next correction being the quadrupole $\sim(\vec{k}\cdot \vec{x})^2$. Unlike the null primordial CMB dipole produced by inflationary perturbations, the lack of cancelation between the gradient term (first term) and the 'potential' term (second term) in the Sachs Wolfe equation Eq.\ref{eq:sachswolfe}, results in the nonzero dipole \cite{turner} of Eq.\ref{eq:inddip}. Thus the predicted dipole and its subsequent bulk flow provide a preferred frame. Since most of the power goes to the dipole, then naturally the alignment plane of the lowest induced multipoles, $\ \le 2$, $\sim(\vec{k}\cdot \vec{x})^l$, is along the gradient of the induced dipole frame.


\begin{figure}[!htbp]
\begin{center}
\raggedleft \centerline{\epsfxsize=3.5in
\epsfbox{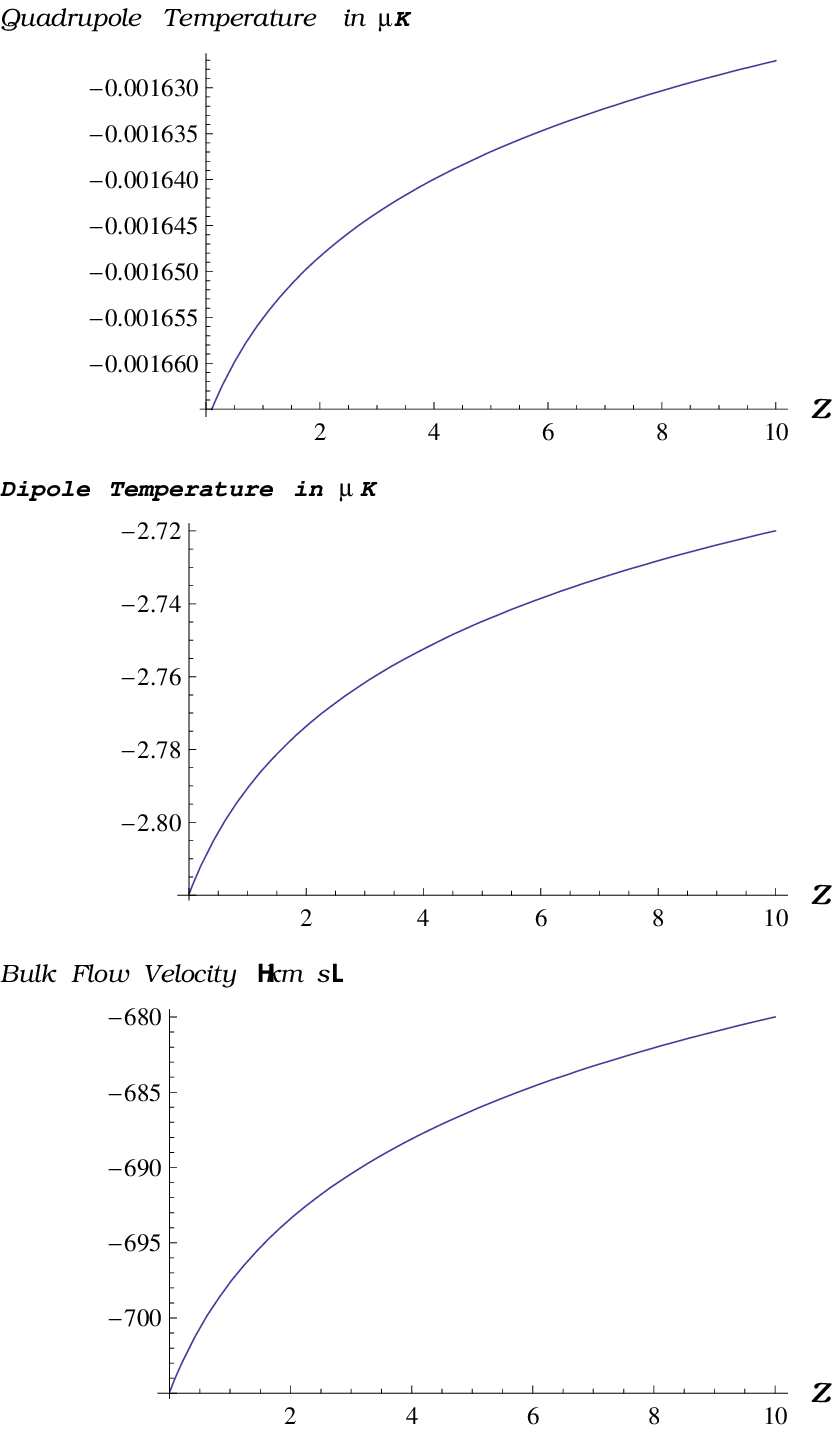}}
\caption{The bulk velocities in km/s generated by entanglement effects and the temperature of the induced quadrupole and dipole in $\mu K$ for representative values of the parameters $b\simeq 3.8 10^{-9}M_{pl}$ and $V_0$ of order the GUT scale.}
\label{fig:bulkvel}
\end{center}
\end{figure}


The natural alignment of the low multipoles for $l \le 2$, with their axes normal to the frame determined by the dominant dipole $(\vec{k}\cdot\vec{x})$, proves useful for discriminating our predictions by making use of polarization maps. The predicted dipole in Eq.\ref{eq:inddip} provides a preferred frame and it gives rise to a bulk flow with velocity Eq.\ref{eq:velocity}. The flow along this dipole preferred frame is relative to the rest frame of the expansion of the universe. The correlation length of such bulk flow, Eq.\ref{eq:velocity}, induced by the noninflationary dipole is of order $r_H$ as we show below. We can now readily estimate the peculiar velocity field from the induced dipole temperature anisotropies through the simple relation

\begin{equation}
\label{eq:velocity}
\beta = \left . \frac{v}{c} \simeq \alpha (\frac{\Delta T}{T})\right |_{\rm dip} \simeq \frac{4\pi}{15} \left (\frac{r_H}{L_1}\right ) \left(\frac{V(\phi) F(b, V(\phi))}{18 M_{\rm Pl}^4}\right )
\end{equation}

 
where the calibration factor $\alpha$ is related to the averaged optical depth of clusters $\tau$. We use a calibration factor of $100km/s$ for each $0.4 \mu K$ anisotropy in converting the dipole temperature of Eq. \ref{eq:inddip} to a peculiar velocity for the bulk flow Eq.\ref{eq:velocity}. This flow has horizon size correlation. Our predicted values for the bulk flow (Eq.\ref{eq:velocity}), the dipole (Eq.\ref{eq:inddip}) and, the quadrupole (Eq.\ref{eq:indquad}), are shown in Fig.1. For $GUT$ scale inflation $L_1$ is of order $900/H$. As can be seen from Fig.1, the induced dipole temperature predicted here is about $3 \mu K$, and the induced quadrupole is suppressed by the factor $H^{-1}/L_{1}$ thus it is about $900-$times less than the dipole.

\begin{figure}[!htbp]
\begin{center}
\raggedleft \centerline{\epsfxsize=3.5in
\epsfbox{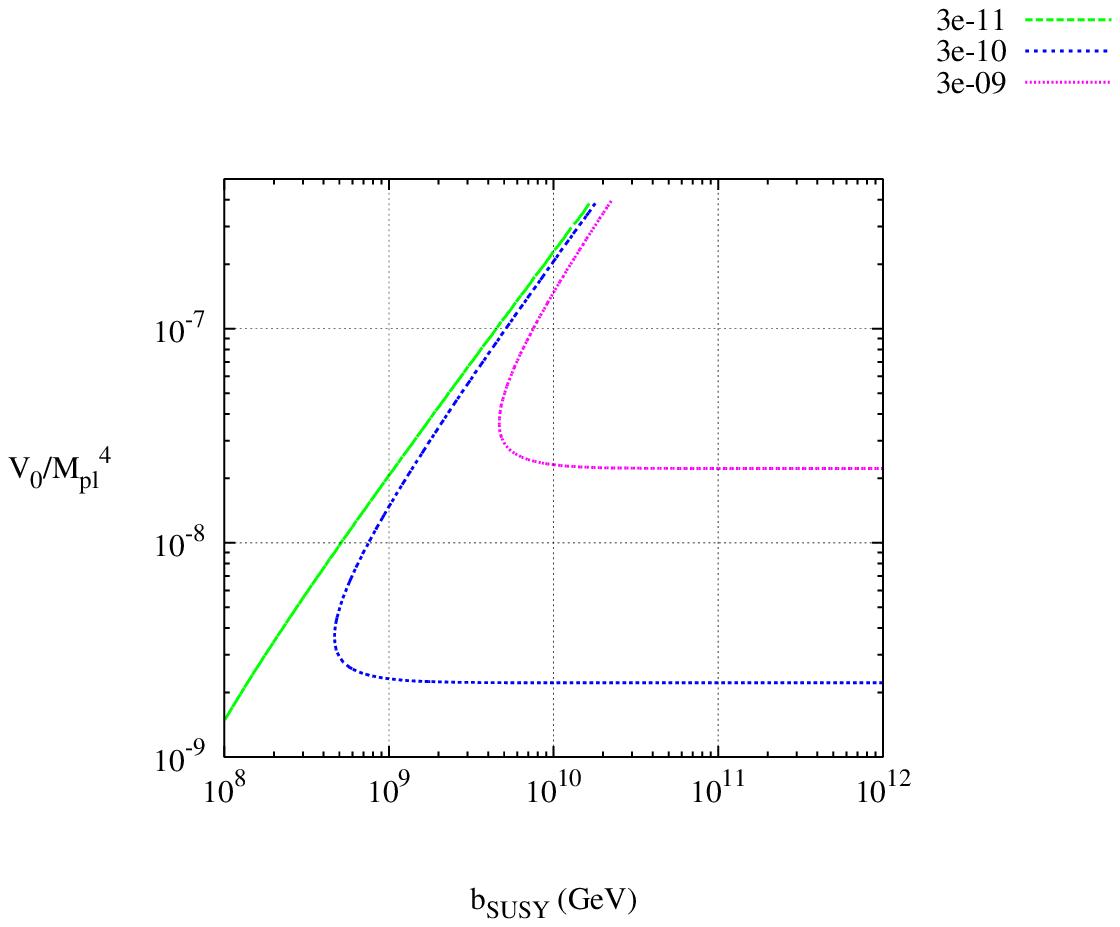}} \caption{Likelihood Contours constraining 'b' as a function of $V(\phi)$ for values of primordial spectrum $P_k$ at $k=0.002
~{\rm Mpc}$ are shown. The value of $\lambda$ is fixed as $\lambda =
0.1$ in this figure. } \label{fig:fig2}
\end{center}
\end{figure}


It is worth emphasizing that the derived expressions for the induced quadrupole and dipole above can {\it not} be assigned different values by a change of parameters. The parameter 'b' corresponding to the initial width of the wavepacket ($SUSY$ breaking scale), is tightly constrained by an upper and lower bound relative to the inflation energy scale $V(\phi)$, derived in \cite{avatars1}. The lower bound is illustrated by the contour plots of Fig.2. These bounds constrain the ratio of the parameters $b/V$, that enter the expression for the quadrupole and octupole, to be a constant, thereby implying that the predictions made by this theory are robust and can be used to test it. 

We show the results of our derivation for the bulk flow driven by the induced dipole in Fig.1. Amazingly, the bulk flow predicted by this theory is of order $650-750 km/s$, a value in agreement with the range observed by \cite{Kash}, (the exact value depends on the calibration factor used) and the predicted dipole is of order $3 \mu K$. The calibration used in Fig.1 corresponds to $100km/s$ flow for a dipole temperature $0.4 \mu K$. In our theory the induced dipole is of noninflationary origin thus its bulk flow provides a preferred direction aligned with it, that moves relative to the expansion frame. This result is in agreement with the 'dark' flow measurements of \cite{Kash}, deduced by comparing the bulk flow field with an all-sky cosmic $X-$ray map which is at rest with the expansion frame. 

An interesting question related to the robustness of our prediction is: would our prediction for the dipole change if we use a different set of values for $(b, V)$, within their allowed range? As can be seen from the likelihood plot of Fig.2 which uses our current limits for the density perturbations to be less than $10^{-5}$, ( $P_k \simeq 10^{-10}$, the blue line in Fig.2), the values of $b$ are constrained by the inflaton energy $V$ to be such that the ratio $b/V$ remains a constant. The 'plateau' in Fig.2 is excluded by the upper bounds on 'b' derived from the slow-roll conditions in \cite{avatars1}. But since all the nonlocal modifications terms involve the ratio of these two scales $b/V$ instead of their values, then these correction terms predict the same strength independent of the chosen values for $b,V$. (It is simple to understand why all the correction terms  induced to: the Friedmann equation $F[b,V]$; to the Newtonian gravitational potential $\delta \Phi$ Eq.\ref{eq:tilt}; to the cumulative coherence length  $L_{1}$ in Eq.\ref{eq:coherencelength}; and for the induced dipole  Eq. \ref{eq:inddip} and quadrupole Eq.\ref{eq:indquad}, are always given as function of the ratio of the two parameters $b$ and $V$. These correction terms are nothing more than 'greybody' factors on the DeSitter temperature of the inflationary domain given by the inflation energy scale. Since in our theory we have two fundamental scale, $'V'$ and $'b'$, then that will give rise to 'greybody' correction terms to the zero order (nearly) DeSitter temperature $'V'$. Greybody corrections are always  a function of the ratio of the two scales since they compare the influence of the 'bath' (the landscape multiverse for our case) onto the system (our Hubble volume) thus the reason why modifications are functions of $b/V$). We had not until now appreciated  in \cite{avatars1,avatars2} the significance of our previous constraints for the constant ratio of $b/V$. To conclude, a constant ratio (see Fig.2 and \cite{avatars1}) between these scales in Eqs.\ref{eq:tilt},\ref{eq:inddip}, implies that our predicted values for the induced dipole, quadrupole and bulk flow, remain the same independently of the energy scale of inflation $V$ in our domain. This conclusion leads to a robust prediction for the 'dark flow' and the dipole since their values are centered around those shown in Fig.1.

Moreover in this framework due to the alignment of the induced low multipoles $\ell \le 1$, the octupole-quadrupole axes normal to their alignment plane, has to be along the same direction as the bulk flow determined by the dominant multipole, the dipole. This leads to the very interesting consequence that the preferred axis addressed in \cite{avatars1,avatars2}, (an observation discussed by many authors \cite{axis}), is not actually a preferred axial symmetry but instead it is a preferred direction defined by the induced dipole. The preferred frame, as demonstrated by the relative flow (the 'dark' flow) of all clusters, including our own local group, with respect to the rest frame defined by the expansion of the universe, is determined by the simple fact that our Hubble domain is embbedded in a much larger environment, the landscape multiverse. The existence of a preferred frame can have major consequences in our interpretation of current data. One such consequence is related to the Doppler shifted photons, due to our 'dark' flow, whereby we would perceive a north-south asymmetry with one side appearing 'hot' and the other 'cold'. The characteristic scale encoding the information about the cumulative nonlocal entanglement with everything beyond the horizon, determines the characteristic wavelength of the induced superhorizon inhomogeneities $L_1$. The superhorizon inhomogeneities and entanglement encoded on CMB and large scale structure (LSS) described here and in \cite{avatars1,avatars2} can account for the list of observed anomalies. Through the predictions made in our theory, which include: the lack of power at the largest scales in temperature anisotropies; the alignement of the lowest multipoles $\ell \le 2$ to the dipole gradient; the existence of a preferred frame given by the induced dark flow of the dipole relative to the universe's expansion rest frame; the north-south asymmetry; and, the unusual high power on the correlation with polarization $C_{l}^{TE}$ \cite{hannestad,axis}, we might be glimpsing at a more coherent picture of cosmology in a multiverse framework.  These testable predictions could place the WMAP\cite{wmap5} and NASA\cite{Kash} findings into context.

\section{Is the Dark Flow a Signal from the Landscape Multiverse?}
Before we discuss tests for discriminating the predictions made by our theory from other mechanisms, let us first briefly review whether other models offer plausible explanation for the bulk flow observed by \cite{Kash}.

The concordance cosmology could in principle produce a bulk flow via unusual superhorizon curvature inhomogeneities, as shown in \cite{turner}. Besides, the problem with $V_{rms}$ mentioned in Sec.1, there are a few more issues with this approach: first, such superhorizon fluctuations have to be tuned to be less than the density perturbations of $O(10^{-5})$ but there is no physical framework or motivation for producing perturbations that are specially tuned to have a specific value, but no less and no more; second, the more serious problem with such a scenario lies in their detection. Even if we accidentally happened to have inflationary superhorizon perturbations, of just the right amount to produce a $700 km/s$ flow, we would still not be able to know of its existence because the spatial hypersurface defined by the isotropy of CMB coincides in this case with the isotropy of expansion, as shown in \cite{turner}. The latter implies that there is no relative flow of one frame with respect to the other. Mathematically the inflationary dipole is canceled by the dipole anistropy of the background Newtonian potential as follows: according to the Sachs-Wolfe effect \cite{sachswolfe,turner} the temperature anisotropies in the direction $\hat{n}$ by an observer today at comoving coordinate $\vec{r}=\vec{0}$ are:
\begin{eqnarray}
\label{eq:sachswolfe}
\frac{\Delta T(\vec{r} = \vec{0};\hat{n})}{T}&=&\frac{1}{2}\left[\sqrt{a} \left(\vec{x}\cdot \nabla\right) \left(\frac{a^2 H^2}{k^2} \frac{\delta \rho}{\rho}\right)\right .\nonumber \\
&+&  \left . \left .\left(\frac{a^2 H^2}{k^2} \frac{\delta \rho}{\rho}\right)\right] \right |_E^R.
\end{eqnarray}
Here $R$, $E$ denote the reception and emission events respectively. The first term could in principle give rise to a dipole anisotropy due to the presence of the gradient operator. However, in the standard inflationary picture the Newtonian potential $\Phi$ is directly related to the induced quadrupole, thus the result of Ref.\cite{turner} that the inflationary induced dipole is {\em canceled} between the first and second term. Furthermore, a cohesive picture of cosmology that is in agreement with present data, requires that besides an accidental 'tilting' of the gravitational potential $\Phi$ by just the right ammount for the flow,  we also have: an 'accidental' alignment of the lowest multipoles along a preferred direction; an anomalously high power on the polarization spectra; but an anomalously low power for the temperature spectra at the largest scales. So many accidental alignments make the likelihood of agreement between a concordance cosmology that doesn't allow for noninflationary perturbations, and the best fit to data, extremely improbable \cite{Kash, turner, wmap5, hannestad,axis}.

This is the essential difference between the standard scenario with our approach; the gravitational potential $\Phi$ gains a contribution from a highly nontrivial second source which is independent of the inflationary curvature perturbations and is a remnant of the superhorizon size nonlocal entanglement from the preinflationary dynamics of our domain in the landscape. For this reason, the induced dipole Eq.\ref{eq:inddip} predicted by our theory of the Initial Conditions, has a different amplitude from the one produced by the second term which ensures that the cancelation does {\em not} occur. As we showed, the expansion frame and the CMB frame do not coincide for this case. Such a net result for the dipole and the quadrupole produces a bulk flow relative to the expansion of the universe frame. The induced quadrupole of Eqn.\ref{eq:sachswolfe} can be used to discriminate among models, as we briefly describe below.

Another possible explanation for the observed dark flow could be provided by scenarios of isocurvature fluctuations \cite{turner} or other models with scalar field perturbations \cite{curvaton}. Isocurvature fluctuations arise from fluctuations of other scalar fields, such as axions or some other PNGB's. Otherwise, the growth and the dynamics of subhorizon and superhorizon isocurvature perturbations is just like that of any other scalar field, including that of inflaton perturbations. Since isocurvature fluctuations are independent of curvature fluctuations, the cancelation between the two dipole contributions does not occur in Eq.\ref{eq:sachswolfe}, for the same reason as the lack of cancelation in our theory. But, in order to get the bulk flow with horizon size correlation among velocities of clusters, superhorizon size fluctuations are needed. Isocurvature modes are highly constrained by current data \cite{wmap5}, to be such that they do not contribute to the energy perturbations in the early universe. Some degree of unnaturalness is required in order to minimize or exclude the subhorizon isocurvature modes and demand that from the onset of inflation these modes were always superhorizon and remain so today. One could imagine there may be a scenario where the subhorizon modes are cuttoff and the tight data constraints are fulfilled by some special isocurvature model. The question thus becomes: can such models give rise to the 'tilted'  universe? Well, surprisingly, due to the expansion of the universe being dominated by dark energy the answer to this question seems to be {\it no} for the following reason: since isocurvature fluctuations are constrained not to contribute to density perturbations, then these fluctuations can grow and excite the background radiation field, only during an axion-dominated universe. In complete analogy with the growth of inflaton perturbations, isocurvature modes are linearly proportional to the scale factor $a[t]$ and can grow {\it only} during a matter dominated universe,. Therefore although their contribution is constrained to be negligible for most of the life of the universe, these modes could still have chance to grow and dominate at late times if the universe becomes axion-matter dominated. If that were the case, isocurvature modes would grow and dominate as a result of which a compensating temperature anisotropy would arise at the dipole level, $\delta T /T |_{dip} \simeq - \delta_{A }/3$. Such a dipole could, in principle, produce the observed dark flow \cite{Kash}.
As we know the universe, is not axion dominated but rather dark energy dominated. The problems is that the growth of perturbations stops during the dark energy domination era. Therefore the isocurvature perturbations can not grow and dominate the energy density of the universe. The condition for exciting the radiation field in order to produce the noninflationary dipole is not fulfilled in a dark energy dominated universe.

An important issue which deserves further investigation, remains: how can we discriminate our predictions here from those of other models? Let us briefly sketch out the following test with polarization maps. We will present this method in detail elsewhere \cite{polarization}. Polarization of CMB is generated through rescattering of the temperature quadrupole. Most of it, ${\bf E_{prim}}$, is generated at the last scattering surface, with a secondary contribution, ${\bf E_{isw}}$, owed to the evolution of the potential $\Phi$ via the ISW effect. For this reason, polarization maps have proven useful for probing dark energy via the ISW effect encoded in ${\bf E_{isw}}$  \cite{asantha}. But, when a bulk motion with peculiar velocity $v$ relative to the rest frame of expansion is present, then there is also a kinematic component of polarization, ${\bf E_k}$, since a quadrupole, proportional to $v^2$, is being generated. The polarization distribution traces out the underlying quadrupole distribution. Thus, for the components ${\bf E_{isw}}$ and ${\bf E_{prim}}$ which trace out the primordial quadrupole produced by inflaton or another scalar, we should expect a random distribution in the polarization skymap. But in our theory, the induced quadrupole proportional to $v^2$ is oriented with respect to a preferred frame determined by the dipole and the transverse velocity field of clusters $v$. Since the correlation in the bulk flow is horizon size then the kinematic component of the polarization map ${\bf E_k}$ that traces out this orientation of the quadrupole will not be random. If we can extrapolate ${\bf E_k}$ from the skymaps we are able to test this theory. This can be done by substracting ${\bf E_{isw}}$ and  ${\bf E_{prim}}$ from the total measured polarization, which removes their random distribution  and allows us to deduce whether ${\bf E_k}$ is random or oriented. Should this mechanism be the right one for addressing the 'dark' flow then the kinetic component of the polarization skymap will not have a random distribution \cite{polarization}. Related to this  test is the fact that the spectral intensity of the kinetic polarization map will be frequency dependent thus provide further information on the predicted alignment of the quadrupole and dipole gradient. Once we have better probes for the ${\bf E_k}$ components, these tests will also be relevant for providing more accurate probes of dark energy via the ISW effect imprinted on the polarization ${\bf E_{isw}}$. Upcoming data from Planck mission will increase the precision of polarization maps and perhaps allow the extrapolation of the kinetic component.

What is surprising is that with these parameter choices, we arrive at bulk flow velocities consistent with recent observations\cite{Kash} and coherent with all the observed anomalies.
Should this theory ultimately be tested to put the findings of \cite{kash} and \cite{wmap5} in context, then such discovery is providing us with a sneak preview of the vast landscape beyond our horizon and, a Copernican Principle extended to the whole universe \cite{tegmark}.


\begin{acknowledgments}
L.~M-H would like to thank A.Kashlinsky for very useful discussions. L.~M-H and R.~H would like to acknowledge support from FQXI foundation. L.~M-H was supported in part by DOE grant DE-FG02-06ER1418 and NSF grant PHY-0553312. R.~H. was supported in part by DOE grant DE-FG03-91-ER40682. 
\end{acknowledgments}

\end{document}